# A quantum feasibility preserving modeling for the min-cut

Ali Abbassi[1,2], Yann Dujardin[2], Eric Gourdin[2], Philippe Lacomme[3*] and Caroline Prodhon[1]

[1] Université de technologie de Troyes, LIST3N, 12 rue Marie Curie - CS 42060 10010 Troyes Cedex, France

[2] Orange Research, 46 Avenue de la république, 92020. Châtillon, France.

[3] Université Clermont-Auvergne, Clermont-Auvergne-INP, LIMOS – UMR CNRS 6158, 1 rue de la Chebarde, 63178 Aubière Cedex, France

**ABSTRACT**

We study the minimum cut problem in weighted undirected graphs using variational quantum algorithms in which only feasible cut configurations are explored. Although minimum cut admits efficient classical solutions, it is a fundamental component of more complex network optimization problems such as multicut and network interdiction. Our objective is to examine quantum models in which feasibility is preserved by the mixer dynamics, without introducing penalty terms in the cost Hamiltonian. We employ a ring-structured *XY* mixer that restricts the quantum evolution to the subspace of valid cut configurations, ensuring that all sampled states correspond to feasible solutions. To address scalability limitations, we suggest an iterative metaheuristic strategy that decomposes large instances into smaller subproblems solved sequentially using the same quantum model. The results obtained using the mixer indicate that the initial probability distribution can be systematically controlled, thereby enabling the development of warm-start techniques within variational quantum-based algorithms.

**KEYWORDS**
QAOA, min-cut, VQA, Quadratic/Linear models, Ring mixer,

# 1 Introduction

1.1. Classical resolution of the Min-Cut

Ensuring the reliability of networked systems frequently necessitates the selective interruption of communication between nodes, subject to budgetary or policy constraints. In the contexts of cybersecurity and network resilience, threat isolation and containment enforcement can often be formulated as the disconnection of designated nodes while minimizing the associated cost. This objective can be captured by the minimum cut (min-cut) problem (Stoer and Wagner, 1997), which, given a weighted graph and two terminal vertices $s$ and $t$, seeks the minimum-weight set of edges whose removal separates $s$ from $t$. The problem is solvable in polynomial time via max-flow duality (Ford and Fulkerson, 1956), and it serves as a foundational primitive for more general separation problems. Extensions such as multiway cuts, minimum multi-cut, bounded-size cuts, and logically constrained cuts are, however, NP-hard (Abbassi *et al*., 2025a, 2025b). In real-world network settings, the scale of the graph and the complexity of the associated constraints often exceed the practical limits of classical approaches, whether based on flow algorithms, linear programming, or path-covering formulations. Consequently, even marginal algorithmic improvements can yield significant practical benefits. This observation motivates the investigation of quantum algorithms, particularly those designed to exploit large combinatorial search spaces. Recent advances in quantum variational heuristics offer a promising alternative to classical techniques, particularly for constrained combinatorial optimization problems characterized by small yet structurally intricate feasible subspaces.

## 1.2. Quantum resolution of the min-cut

The minimum cut problem has been investigated under two distinct paradigms in the quantum computing literature. The first paradigm, developed primarily by (Apers and Lee, 2020) and extended by (Apers *et al.*, 2021), focuses on establishing quantum query complexity bounds for the problem. These works analyze different graph access models—such as adjacency matrix and adjacency list representations—and exploit structural properties of the minimum cut problem to derive complexity improvements over classical algorithms. A key result is the proof that there exists a quantum algorithm that computes the minimum $s-t$ cut value with high probability using

$$\tilde{O}(\sqrt{m} n^{\frac{5}{6}} W^{\frac{1}{3}})$$

queries, up to polylogarithmic factors, where $m$ denotes the number of edges, $n$ the number of vertices, and $W$ the maximum edge weight. This query complexity is strictly sublinear in $m$, in contrast to classical randomized algorithms, for which a lower bound of $\Omega(m)$ queries is required.

A second, complementary paradigm addresses the minimum cut problem from an optimization and implementation-oriented perspective. In particular, (Krauss *et al.*, 2020) tackle the maximum flow problem by leveraging the Max-Flow Min-Cut Theorem and formulating the corresponding minimum cut problem as a quadratic unconstrained binary optimization (QUBO) model. In their approach, penalty terms are introduced to enforce the source and sink constraints, while the objective function minimizes the total capacity of edges separating the vertex sets $S$ and $T$. The remainder of their work is devoted to analyzing the resource requirements of this formulation, including qubit counts, the number of quadratic terms, embedding overhead, and additional constraints arising from the use of D-Wave's quantum annealing hardware.

## 1.3. Variational approaches

Optimization problems can be represented in multiple, formally equivalent ways (choice of variables, objective, constraints, penalties, encodings). However, the chosen formulation strongly affects which properties are made explicit (feasible-set structure, locality, landscape geometry) and therefore affects how efficiently a given solver—classical or quantum—can explore the solution space. This "modeling matters" standpoint is emphasized in the broader quantum-optimization literature (Abbas *et al.*, 2024)

Within this perspective, it is natural to distinguish several variational "settings" for VQA-type algorithms, even when targeting the same underlying min-cut objective.

(1) Setting I — Vanilla QAOA with expectation-value training (baseline)

We encode the (min-)cut objective as a diagonal cost Hamiltonian $H_C$ (e.g., via an Ising/QUBO mapping). The standard depth-p QAOA ansatz alternates a problem unitary generated by $H_C$ and a transverse-field mixing unitary generated by $H_M = \Sigma_j X_j$, starting from the uniform superposition $|\psi_0\rangle = |+\rangle^{\otimes n}$ (Farhi, *et al.*, 2014):

$$|\psi(\gamma,\beta)\rangle = \prod_{\{\ell=1\}}^{p} e^{-i\beta_\ell H_M} e^{-i\gamma_\ell H_C} |+\rangle^{\otimes n}.$$

Because $|+\rangle^{\otimes n}$ assigns equal amplitude to all bitstrings, the initial state "spreads mass" across the entire search space; successive alternations are intended to shape interference so that probability concentrates around low-cost (high-quality) solutions.

The conventional training objective is the energy expectation (estimated from measurement samples):
$$F_{E(\gamma,\beta)} := \langle \psi(\gamma,\beta)|H_C|\psi(\gamma,\beta)\rangle.$$

In a black-box (setting parameters γ,β) are updated by a classical optimizer. A common choice is COBYLA, which is a local, derivative-free (gradient-free) method based on linear approximations, and is therefore compatible with noisy objectives estimated from finite sampling (Powell, 2007).

**(2) Setting II — QAOA as a Quantum Alternating Operator Ansatz (feasibility-preserving mixers)**

A second, structurally different setting arises when the problem formulation induces a nontrivial feasible set $F \subset \{0,1\}^n$ (hard constraints). In such cases, the transverse-field mixer $H_M = \Sigma_j X_j$, described in setting I, generally moves the probability distribution over the candidate solutions outside F, so feasibility must be enforced indirectly (penalties, postselection, or repair). The Quantum Alternating Operator Ansatz viewpoint replaces the generic mixer with a feasibility-preserving mixing operator tailored to the constraint structure. (Hadfield *et al*, 2018). Let $H_F := span\{|z\rangle : z \in F\}$ be the feasible subspace. A feasibility-preserving mixer is a parameterized unitary $U_M(\beta)$ such that:

$$U_M(\beta) H_F \subseteq H_F,$$

so that starting from $|\psi_0\rangle \in H_F$, all intermediate states remain feasible. The alternating circuit becomes:

$$|\psi(\gamma,\beta)\rangle = \prod_{\{\ell=1\}}^{p} U_M(\beta_\ell) e^{-i\gamma_\ell H_C} |\psi_0\rangle,$$

where $|\psi_0\rangle$ is a feasible initial state (often a simple feasible configuration or a prepared feasible superposition).

**(3) Setting III — Same circuit family, alternative training objectives (tail-focused learning such as CVaR)**

For combinatorial optimization, the end goal is typically not "a low expectation value" per se, but rather to sample very good bitstrings with high probability. This opens a second degree of freedom: keep the same circuit family $U(\gamma,\beta)$, but modify the scalar objective used to learn $(\gamma, \beta)$ from the same measurement outcomes. A canonical example is the Conditional Value-at-Risk (CVaR), introduced for variational quantum optimization as a sample-aggregation objective that emphasizes good outcomes (Panagiotis *et al.*, 2020).

Let the $K$ measured costs be $E_1, \dots, E_K$ and let $E_1 \leq \dots \leq E_K$ be the sorted list. For $\alpha \in (0,1]$, define

$$m = \lceil \alpha K \rceil \text{ and } CVaR_{\alpha(\gamma,\beta)} := \left(\frac{1}{m}\right) \Sigma_{\{j=1\}}^{m} E_j.$$

Key mathematical distinctions relative to the expectation objective:

- Expectation objective ($\alpha = 1$): averages all samples equally,
  $F_E = \frac{1}{K} \Sigma_{\{k=1\}}^{K} E_k$ (in Monte Carlo form).
- CVaR objective ($\alpha < 1$): averages only the lowest-energy α-fraction (the "best tail"),
  $CVaR_\alpha = \frac{1}{m} \Sigma_{\{j=1\}}^{m} E_j$, which explicitly biases training toward increasing probability mass on high-quality samples.

Practical note on "3rd quartile": if one informally says "third quartile" (0.75), then $CVaR_{0.75}$ averages the best 75% (lowest 75%) of observed energies for a minimization problem; it is not the upper quartile. By varying α and/or composing tail-based objectives with min-like statistics (e.g., soft-min/log-sum-exp), one obtains a family of learning criteria ranging from mean-like (stable, but less selective) to min-like (highly selective, but higher-variance under finite shots). This shift paradigm lies within the framework of the variational quantum algorithm as we seek to find by the end of the optimization a close upper bound to the ground state. The set of algorithms belonging to this class of methods is divided into four categories.

The set of algorithms belonging to this class of methods is divided into four categories.

- **Vanilla QAOA**
  - Unconstrained combinatorial optimization problems (Max Cut like)
  - $H_p$ and $H_M = \Sigma_j X_j$
- **Quantum Alternating Operator Anzats**
  - Unconstrained combinatorial optimization problems (Graph Coloring)
  - $H_p$ and feasibility preserving mixer $H_M$ such as the *XY mixer*
- **VQE: Variational Quantum Eigensolver**
  - Initially designed for Quantum Chemistry
- **Custom Variational Quantum Algorithm**
  - A custom choice of the cost function to minimize

Empirical studies have consistently validated the effectiveness of structure-aligned mixers in variational quantum algorithms. Notable contributions by (Bourreau *et al.,* 2022), (Govia *et al.*, 2021), and (Tsvelikhovskiy *et al.*, 2024) demonstrate that such mixers not only improve convergence behavior but also significantly reduce the circuit depth required to obtain high-quality solutions. In particular, et al. introduce a symmetry-based framework for constructing QAOA mixers, leveraging group-theoretic properties of the objective function to guide both the design and parameterization of quantum operators.

In this work, we adopt the Variational Quantum Algorithm (VQA) framework to develop a constraint-preserving quantum approach for solving the minimum cut problem. Our circuit architecture incorporates a ring-based XY mixer, which ensures that the quantum evolution remains confined to valid interdiction configurations, thereby restricting the modelling to the feasible subspace. The cost Hamiltonian is derived from a path-based representation of the network, enabling a direct encoding of arc-level interdictions without introducing penalty terms or slack variables. This methodology aligns with a broader trend in feasibility-preserving variational algorithms, wherein problem-specific structure — exploited both in Hamiltonian construction and mixer design — is leveraged to improve performance and scalability in constrained optimization tasks.

The remainder of the paper is organized as follows. Section 2 formulates the minimum cut problem and presents the corresponding classical path-based model. Section 3 details the construction of the cost and mixer Hamiltonians, introduces an iterative search-based solution strategy for large-scale graphs, and discusses warm-start techniques applicable within this framework. Section 4 presents numerical simulation results, and Section 5 concludes with a discussion of the findings and outlines directions for future research.

## 2 Problem Statement

### 2.1. Description

Let $G = (V, E)$ be a directed, weighted graph, with source node $s \in V$, and target node $t \in V$. The minimum cut problem seeks a minimal cost subset of edges $E' \subseteq E$ whose removal disrupt all $s - t$ paths.

Each edge $e_k$ in the graph is associated with a positive weight. This setup naturally arises in applications such as threat interdiction or resilience modeling, where specific communication routes must be disrupted under resource constraints. While the problem is solvable in polynomial time, real-world scenarios often introduce challenges when integrating solutions into large, dynamic systems or when rapid reconfiguration is required. Figure 1 depicts a sample weighted graph, where node 4 is referred to as the source and node 2 as the sink. Each arc, labelled in a square box from 1 to 10, is assigned a positive weight which represents its cost.

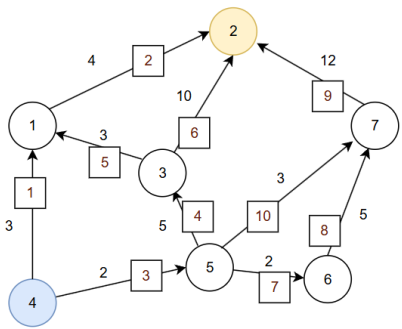
Figure 1: An example of a weighted graph

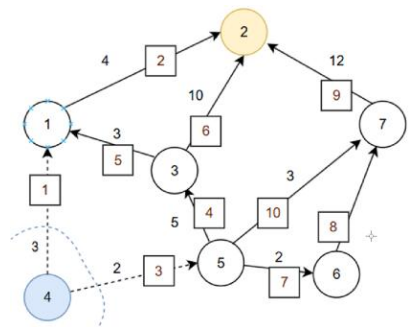
Figure 2: Optimal solution

A solution is defined as a set of arcs such that no path exists from the source to the sink. The cost of the solution is the sum of the costs associated with the removed arcs. For instance, the optimal solution for the graph depicted in Figure 1 is represented in Figure 2 and involves removing the arcs (4,1) labelled 1 and the arc (4,5) labelled 3, resulting in a total cost of $3 + 2 = 5$.

We first introduce the classical path-based linear formulation, where the goal is to disconnect all known source-to-sink paths by removing a minimum-cost set of arcs. In transitioning to a quantum setting, we reinterpret this formulation in terms of quantum states and Hamiltonians. The key idea is to encode both the objective and the feasibility constraints directly into the structure of the quantum circuit, using cost and mixer operators that mimic the logic of the classical model. This requires careful mapping of arc removals to spin variables and the design of unitaries that restrict evolution to valid interdiction configurations.

## 2.2. Linear formulation

We present in the following the formulation of the min cut problem, with the notations and binary decision variables defined in Table 1.

**Table 1: Modelling notations**

| | |
|---|---|
| **Sets and parameters** | |
| $V$ | Set of vertices |
| $E$ | Set of arcs |
| $n = |E|$ | Number of arcs in the graph |
| $S$ | Source node |
| $T$ | Sink node |
| $P = \pi^i$ | Set of $s - t$ paths |
| $np = |P|$ | Number of paths in the graph |
| $\pi^i$ | path number $i$ s.t. $\pi^i = [e_1 = s, e_2, \ldots e_{n_i} = t]$ |
| $A_i^j$ | $a_i^j \in \{0,1\}$, values 1 if arc $i$ appears in path j, 0 otherwise. |
| $n_i$ | the number of arcs in the path $\pi^i$ |
| $W_i$ | weight of arc $i$ with $i = 1 \ldots n$ |
| **Variables for the linear formulation** | |
| $x_i$ | $x_i \in \{0,1\}$, 1 if the arc is removed |
| **Variables for the quantum formulation** | |
| $y_k$ | $y_k \in \{0,1\}$ |
| $x_j^i$ | $x_j^i = 1$ indicates that the arc in position $j$ of path $\pi^i$ is removed |

This formulation assumes an explicit enumeration of all source-to-sink paths and ensures that at least one arc is removed on each path. Though the number of constraints may grow exponentially in the worst case, this model maps naturally to quantum encodings via path-wise binary registers.

$$\text{Minimize} \sum_{i=1}^{n} x_i \cdot w_i$$

$$subject\ to \quad \sum_{i=1}^{n_j} A_i^j \cdot x_i \geq 1 \quad \forall j = 1..np$$

This model guarantees that for each path $\pi^j \in P$ at least one arc is removed. The objective function minimizes the total removal cost. This formulation will serve as the basis for the cost Hamiltonian in the quantum version of the problem. Although not used in our quantum construction.

# 3 Proposition

## 3.1. Cost Hamiltonian Construction

Let $G = (V, E)$ be a directed, weighted graph, with a designated source node $s \in V$, a target node $t \in V$, and a collection $P = \{\pi^1, \ldots, \pi^{np}\}$ of known source to sink paths. Each path $\pi^i$ is defined as an ordered sequence of arcs $\pi^i = [e_1^i, \ldots, e_{n_i}^i]$ where $n_i$ denotes the length of the path.

We model each source-to-sink path as a disjoint constraint, with the assumption that removing a single arc is sufficient to disrupt the corresponding path. This implies that, for each path, exactly one arc must be marked for removal. However, since the underlying graph may contain overlapping paths, some arcs may appear in multiple paths and be selected for removal more than once at the path level. To prevent overcounting in the objective function, we ensure that each arc's contribution to the total cost is counted only once, regardless of how many paths select it for interdiction.

To encode interdiction, we associate a binary register $\vec{x} = [x_j^i]$ to the path level: $x_j^i = 1$ indicates that the arc in position $j$ of path $\pi^i$ is removed. Our goal is to determine whether an edge $e_k \in E$ is removed across all paths. This is captured by auxiliary binary variables $y_k \in \{0,1\}$, where:

$$y_k = 1 - \prod_{\substack{i=1 \\ j=1 \\ \pi^i(j)=k}}^{\substack{i=np \\ j=|\pi^i|}} (1 - x_j^i)$$

This constraint ensures that $y_k = 1$ if we have one or more paths where the arc $k$ is used (i.e. for one path $i$ we have one position $j$ where $\pi^i(j) = k$). The total cost is then:

$$\min \sum_{k=1}^{n} w_k \cdot y_k$$

**Example: We illustrate on the previous figure:**

Let us assume we have path 1 noted $\pi^1 = [1,2]$ and path 2 such as $\pi^2 = [3,4,5,2]$, represented respectively in black and red in Figure 3.

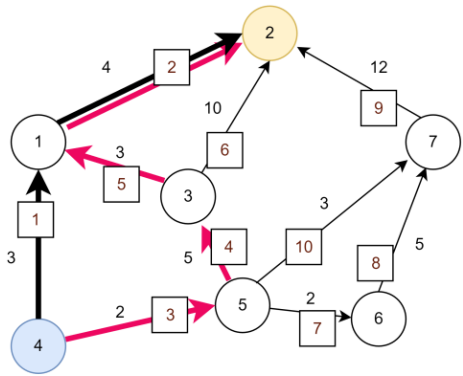

Figure 3: Two paths into the graph

If we have $x_1^1 = 1$, it means that the arc (4,1) labelled arc 1 is removed and since it is not used in path 2, we obtain:

$$y_1 = 1 - \prod_{\substack{i=1 \\ j=1 \\ \pi^i(j)=k}}^{\substack{i=np \\ j=|\pi^i|}} (1 - x_j^i) = 1 - 0 = 1$$

If we have $x_2^2 = 1$, it means that the arc (5,3) labelled arc 4 is removed in path 2 only (and not used in path 1) we obtain:

$$y_4 = 1 - \prod_{\substack{i=1 \\ j=1 \\ \pi^i(j)=k}}^{\substack{i=np \\ j=|\pi^i|}} (1 - x_j^i) = 1 - 0 = 1$$

If we have $x_2^1 = 1$ it means that the arc (1,2) labelled arc 2 is removed in path 1 and simultaneously in path 2 ($x_4^2 = 1$) we obtain:

$$y_2 = 1 - \prod_{\substack{i=1 \\ j=1 \\ \pi^i(j)=k}}^{\substack{i=np \\ j=|\pi^i|}} (1 - x_j^i) = 1 - (1 - x_2^1)(1 - x_4^2) = 1 - 0 \times 0 = 1$$

The objective function consists in the sum of the arc removed ($y_k = 1$):

$$\min \sum_{k=1}^{n} w_k \cdot y_k$$

A more compact formulation is:

$$\min \sum_{k=1}^{n} w_k \cdot \left( 1 - \prod_{\substack{i=1 \\ j=1 \\ \pi^i(j)=e_k}}^{\substack{i=np \\ j=|\pi^i|}} (1 - x_j^i) \right)$$

We assume now that we know the 5 paths the graph of Figure 3 to link the source and the sink. Let the initial state be composed of a set of arcs removed on each path as follows:
- Arc 4-1 is removed in the path $4 - 1 - 2$
- Arc 3-1 is removed in the path $4 - 5 - 3 - 1 - 2$
- Arc 4-5 is removed in the path $4 - 5 - 3 - 2$
- Arc 4-5 is removed in the path $4 - 5 - 6 - 7 - 2$
- Arc 5-7 is removed in the path $4 - 5 - 7 - 2$

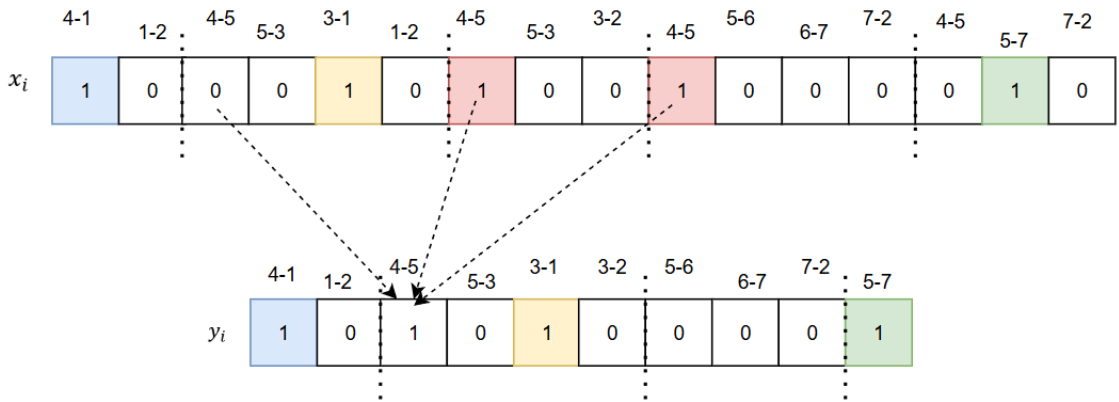

Figure 4: Arc interdiction vector $x$ and its effect on path feasibility via vector $y$

Thus, we have, $x_1^1 = 1$ i.e. the arc number 1 in path 1 is removed, $x_3^2 = 1$ i.e. the arc number 3 in the path 2 is removed and so on. The binary variable $y_k$ depends on $x_j^i$, $y_k = 1$ if the arc $k$ is removed in the graph. For example, $y_1 = 1$ meaning that arc $4 - 1$ is removed and $y_3 = 1$ meaning that arc $4 - 5$ is removed in the example introduced in Figure 4. The mixer is applied to the paths defined by the variables $x_j^i$. For example, path number 2, which includes the arcs 4–5–3–1–2, is represented by positions 3, 4, 5, and 7 in

Figure 4. The table of $y_i$ consists of binary values and is used to compute the cost of the solution, as each removed arc appears exactly once thus preserving the Hamming weight.

The encoding strategy relies on the principle that disabling any single arc along a path suffices to disrupt connectivity from source to sink. To enforce this, the formulation imposes the removal of exactly one arc per path but several arc can be removed to a path due to overlapping path.

The above construction induces high-order interaction terms, since each variable $y_k$ is expressed as a product over all path-level variables associated with corresponding edge. In the corresponding quantum Hamiltonian, this leads to multi-qubit operators whose locality increases with the number of paths containing the same edge. Alternative formulations could be considered to control this nonlocality, for instance by introducing auxiliary qubits to linearize the logical relations, or by implementing explicit counting and comparison circuits through quantum adders. Such approaches would reduce the degree of the interaction terms at the cost of additional qubit resources and circuit depth.

This ensures consistent and non-redundant cost evaluation. Although the quantum circuit selects one arc per path to satisfy feasibility constraints, the resulting solution may involve multiple arcs removed along the same physical path, due to the convergence of removal decisions across different paths. This is a natural outcome of the non-disjoint structure of the path set. The responsibility for exploring feasible interdiction combinations—each consisting of one removed arc per path—lies with the mixer Hamiltonian. It ensures that the quantum state remains confined to valid configurations while allowing the optimization process to identify cost-efficient solutions under the defined constraints.

By expending and reparametrizing $x_j^i \in \{0,1\}$ to Pauli-Z space via the transformation $\frac{1}{2}(Id - Z_j^i)$, we obtain a cost Hamiltonian $H_p$ expressed in terms of tensor products of Pauli-Z operators. This yields a nonlinear, yet efficiently evaluable operator whose structure mirrors classical logic. For example, for the instance shown in figure 3, we have $|E| = 10, |V| = 7$, and the resulting Hamiltonian cost takes the form:

$$H_p = \sum_{k=1}^{n} w_k \cdot \left( 1 - \prod_{\substack{i=1 \\ j=1 \\ \pi^i(j)=e_k}}^{\substack{i=np \\ j=|\pi^i|}} \left( \frac{Id + Z_j^i}{2} \right) \right)$$

This formulation avoids additional qubits and is directly implementable using Pauli gates, making it particularly well-suited to NISQ-era devices.

### 3.2. XY Ring Mixer Hamiltonian formulation

If we consider a qubit-string of $n$ qubits then

$$H_M = \sum_{i=0}^{n} \frac{1}{2} (X_i \cdot X_{i+1} + Y_i \cdot Y_{i+1})$$

with $n + 1 = 0$

It defines one Hamiltonian such that $K$ hamming weight of the qubit string is conserved.

Let each path $\pi^i$ be encoded via a one-hot qubit register $\vec{x_j^i} = [x_1^i, \dots, x_{n_i}^i]$ where feasibility requires:

$$\sum_{j=1}^{n_i} x_j^i = 1$$

To preserve this constraint, the mixer for path $\pi^i$ can be defined as:

$$H_M^{(i)} = \sum_{j=1}^{n_i-1} \frac{1}{2}(X_j \cdot X_{j+1} + Y_j \cdot Y_{j+1})$$

This is the **XY chain** with periodic or open boundary conditions, depending on the path's length. The total mixer Hamiltonian is:

$$H_D = \sum_{i=1}^{np} H_M^{(i)}$$

This operator preserves the structure by only permuting amplitude among valid basis states, allowing us to hop from a feasible state to another during the exploration phase of the algorithm.

The last ingredient of the proposition is quantum state preparation. The idea is to determine an educated guess about the initial state we prepare such that we bridge the gap toward the best solution.

### 3.3. Preparing an initial state: Warm-start

Since we have numerous classical approaches that can provide high quality solutions in short computation time, it is possible to define one $|\psi_0\rangle$ where a large part of the distribution is concentrated on such quality solution. Such use of feasible initial states is referred to as a warm-start approach meaning that the initial wave is defined to favor convergence of the method or to give some very specific shape to the initial probability distribution of the quantum state.

The mixer for path $\pi^i$ can be defined as:

$$H_M^{(i)} = \sum_{j=1}^{n_i} \frac{1}{2}(X_j \cdot X_{j+1} + Y_j \cdot Y_{j+1}) = \sum_{j=1}^{n_i-1} H_{M,j}^{(i)}$$

where $H_{M,j}^{(i)}$ refers to as the swap of qubit $j$ and qubit $(j + 1) \bmod n_i$ with the classical convention of (Hadfield *et al.*, 2019).

So

$$e^{-i.\vec{\alpha}.H_M^{(i)}} \cong \prod_j e^{-i.\alpha_i.H_{M,j}^{(i)}} = \prod_j \left(\cos\alpha_i . Id - i.\sin\alpha_i . H_{M,j}^{(i)}\right)$$

And

$$e^{-i.\alpha_i.H_M^{(i)}} . |\psi_0\rangle = \left[\prod_j \left(\cos\alpha_i . Id - i.\sin\alpha_i . H_{M,j}^{(i)}\right)\right] . |\psi_0\rangle$$

**Example**

Then if we consider $n_i = 4$ and $|\psi_0\rangle = |0100\rangle$

$$e^{-i.\alpha_i.H_M^{(i)}} . |\psi_0\rangle \cong e^{-i.\alpha_i.H_{M,3}^{(i)}} e^{-i.\alpha_i.H_{M,2}^{(i)}} e^{-i.\alpha_i.H_{M,1}^{(i)}} e^{-i.\alpha_i.H_{M,0}^{(i)}} . |0100\rangle$$

it is possible to have the functional explicit wave definition.

□

The choice of using the same angle $\alpha_i$ to define the probability between any qubit and its successor is an arbitrary assumption made during the modeling process. One could instead define, for the $n_i$ qubits, an angle vector $\vec{\alpha}_i = (x, y, z, t)$, which would lead to the following distribution:

$$e^{-i.\vec{\alpha}_i.H_M^{(i)}} . |\psi_0\rangle \cong \cos t . \cos z . \cos y . \cos x . |0100\rangle - i.\cos t . \cos z . \cos y . \sin x . |1000\rangle$$
$$- i.\cos t . \cos z . \sin y . \cos x . |0010\rangle - \cos t . \cos z . \sin y . \sin x . |1000\rangle - i$$
$$\ldots$$
$$+ i.\sin t \sin z . \sin y \cos x . |0001\rangle + \sin t \sin z . \sin y \sin x . |1000\rangle$$

By choosing the angles $(x, y, z, t)$, one can define an arbitrary probability distribution. For instance, by choosing $\left(0, \frac{\pi}{2}, 0, 0\right)$, one obtains:

$$e^{-i.\vec{\alpha}_i.H_M^{(i)}}.|\psi_0\rangle \cong |0010\rangle$$

Using

$$\left(0.1, \frac{\pi}{2} - 0.1, 0.1, 0.1\right)$$

The term

$$-i.\cos t.\cos z.\sin y.\cos x = -i.\cos 0.1.\cos 0.1.\sin\left(\frac{\pi}{2} - .1\right).\cos 0.1$$

defining a probability of 0.96 to $|0010\rangle$

Similarly, one can consider the term $\cos t.\cos z.\cos y.\cos x.|0100\rangle$ and compute

$$\cos t.\cos z.\cos y.\cos x = \cos 0.1.\cos 0.1.\cos\left(\frac{\pi}{2} - 0.1\right).\cos 0.1 = 0.098$$

defining a probability of 0.096 for the state $|0100\rangle$.

### 3.4. Iterative QAOA based resolution

The former modeling approach is based on the explicit enumeration of paths, which are used to define the global optimization model and subsequently transformed into a Hamiltonian formulation. For very large graphs, an iterative scheme can be employed in which the set of paths is explored progressively. The underlying principles are presented in Algorithm 1 and rely on the selection of an initial set of paths in the original graph. It is reasonable to consider a set of paths with a high degree of arc overlap, as this is likely to facilitate the derivation of a solution that requires the removal of only a small number of arcs to eliminate these paths. It should also be noted that the number of initially selected paths is an important parameter.

The main loop consists of: (i) generating the Hamiltonian that models the problem using the selected paths; (ii) performing a quantum optimization to obtain a solution, i.e., a set of arcs whose removal breaks these paths; (iii) effectively removing these arcs from the graph; and (iv) iterating the process until no paths remain. This type of algorithm avoids the costly—if not infeasible—enumeration that may arise for large-scale graphs within a reasonable time frame.

---

**Algorithm 1.** Iterative resolution with QAOA

**Input parameters:**
  $G = (E, V)$
**Output parameters:**
  $S$ : solution found (set of arcs to remove from the graph)
**Begin**
  Compute one initial set of paths: $P = \{set\ of\ path\}$
  $S = \{\}$
  **While** $(|E| \neq 0)$ **do**
   Generate the Hamiltonian related to the set of paths in $P$
   $PS$ = **Call** QAOA to have a quantum resolution of the problem
   $E = E - PS$
   $S = S \cup PS$
  **End While**
**End**

---

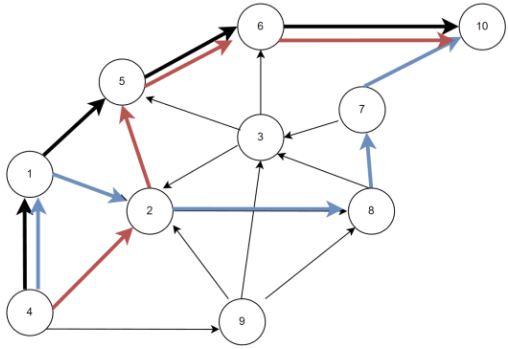

Figure 5: Selection of 3 initial paths

The principle can be illustrated using the graph given in Figure 5. Three paths (shown in black, red, and blue) can be considered among all possible paths i.e.

$$P = \{(4,1,5,6,10), (4,1,2,8,7,10), (4,2,5,6,10)\}$$

The first optimization step then leads to the selection of arc (4,1), which eliminates the black and blue paths, and arc (4,2), which eliminates the red path. Both arcs are thus suppressed of the graph as shown in Figure 6.

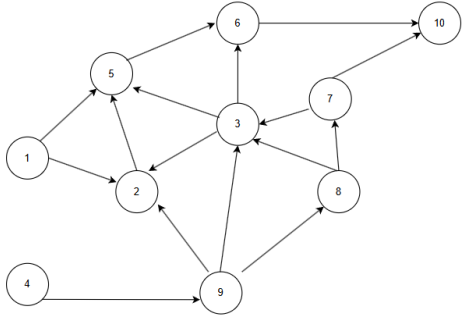

Figure 6: Graph after suppression of (4,1) and (4,2)

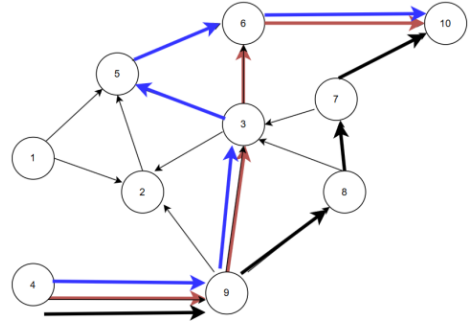

Figure 7: Three paths in the new graph

In this updated graph, it is then possible to select a subset of the remaining paths, for example the three paths shown in Figure 7, and the iterative process can continue.

## 4 Numerical Validation

### 4.1. Ring mixer

Three numerical experiments are performed on a graph with 10 weighted arcs, using quantum circuits of 16 qubits implemented in Qiskit and simulated with the AerSimulator backend.

An initial feasible cut configuration is prepared by applying Pauli-$X$ gates to selected qubits, thereby encoding a high-quality classical solution in the computational basis. A parameterized circuit composed of alternating cost and mixing layers is then applied, with feasibility preserved by the joint design of $H_D$ and $H_P$.

The initial probability distribution is determined by the initial solution and the angles assigned to the ring mixer. Table 2 illustrates two representative settings for the 7-vertex instance, in which the qubits $\{0, 9, 13, 7, 3\}$ are assigned mixer angles of 0.5, and $\{1, 10, 13, 7, 3\}$ are assigned mixer angles of 0.25, respectively.

We observe that the probability distribution is fully determined by the choice of the initial state and the mixing angle. This choice defines a warm start, in the sense that it allows explicit control over the initial probability distribution. Such a warm start enables the exploitation of high-quality solutions obtained using classical (non-quantum) approaches.

Table 2: Probability distribution left shifted for a given instance

| Experiment 1 | | Experiment 2 | |
| --- | --- | --- | --- |
| Cost | Probability distribution defined by the mixer | Cost | Probability distribution defined by the mixer |
| 6.0 | 6.7 | 6.0 | 0,2 |
| 8.0 | 2.1 | 7.0 | 0,6 |
| 9.0 | 7.0 | 8.0 | 1 |
| 10.0 | 0.7 | 9.0 | 0,2 |
| 11.0 | 4.9 | 10.0 | 2.6 |
| 12.0 | 0.7 | 11.0 | 3.0 |
| 13.0 | 1.3 | 12.0 | 14.90 |
| 14.0 | 2.5 | 13.0 | 11.80 |
| 15.0 | 0.3 | 14.0 | 2.3 |
| 16.0 | 12.7 | 15.0 | 6.2 |
| 17.0 | 0.6 | 16.0 | 5.5 |
| 18.0 | 10.2 | 17.0 | 3.3 |
| 19.0 | 10.5 | 18.0 | 1.7 |
| 20.0 | 1.8 | 19.0 | 1.9 |
| 21.0 | 5.7 | 20.0 | 3.1 |
| 22.0 | 3.3 | 21.0 | 3.2 |
| 23.0 | 4.6 | 22.0 | 9.9 |
| 24.0 | 1.7 | 23.0 | 8.6 |
| 25.0 | 0.6 | 24.0 | 3.8 |
| 26.0 | 1.7 | 25.0 | 3.4 |
| 27.0 | 0.1 | 26.0 | 3.5 |
| 28.0 | 12.6 | 27.0 | 0.8 |
| 29.0 | 0.3 | 28.0 | 0.7 |
| 30.0 | 0.4 | 29.0 | 1.4 |
| 31.0 | 5.1 | 30.0 | 0.7 |
| 32.0 | 0.3 | 31.0 | 1.2 |
| 33.0 | 0.8 | 32.0 | 2.6 |
| 34.0 | 0.4 | 33.0 | 1.4 |
| | | 34.0 | 0.2 |
| | | 35.0 | 0.3 |
| | | 36.0 | 0.1 |

For example, let us consider the initialization where the qubits $\{1, 10, 13, 7, 3\}$ are assigned to $|1\rangle$ which is the second example of table 3.

By assigning 1 to qubit 1, we defined a solution where (figure 4):
- the arc $1 - 2$ is removed in the graph,
- the arc $5 - 6$ is removed in the graph,
- the arc $4 - 5$ is removed in the graph,
- the arc $4 - 5$ is removed in the graph.

This solution is related to the cost 13 which is the cost of the removed arcs, and by defining relatively small mixer angle in $e^{-i.\vec{\alpha}_i.H_M^{(i)}}$, it leads to a distribution concentrated on the solution of cost 13 meeting the theoritical consideration introduced in the previous section.

The analysis of the convergence of the variational algorithm, as well as the definition of objective functions appropriate for achieving the desired accuracy and solution quality, are beyond the scope of this work. Instead, in the next subsection, we highlight the complexity of this task by discussing the challenges associated with the optimization phase.

### 4.2. Example of VQA space investigation

Let $f : \{0,1\}^n \rightarrow \mathbb{R}$ denote a classical cost function associated with the combinatorial optimization problem under consideration. The bitstring distribution $p_\theta$ induces a probability distribution over cost values through the pushforward of $p_\theta$ by $C$ defined by

$$q_{\theta(c)} = \sum_{\{x \in \{0,1\}^n : C(x) = c\}} p_{\theta(x)}$$

This cost distribution $q_\theta$ is the central object governing the sampling behavior of the quantum circuit with respect to the optimization problem and in practice, $q_\theta$ estimated from a finite number $N$ of circuit executions, yielding an empirical estimator $\hat{q}_\theta^{(N)}$. To investigate the search space of solutions, each solution defined by a vector $\theta$, it is possible to consider any relevant optimization methods. Here we consider only two classical generic methods: a population based one and derivated-free optimization that have been chosen to give some basic examples of search space exploration. The population-based genetic algorithm is implemented using the pygad library, without any embedded local-search or hybrid refinement mechanism and the derivative-free optimization method, namely Powell's algorithm as implemented in SciPy, which performs sequential line minimizations and is characterized by local convergence behavior and strong sensitivity to initialization.

Rather than optimizing only the expected cost $E_{q_\theta}[C]$, we define an objective function that explicitly depends on distributional properties of $q_\theta$. Specifically, we introduce

$$f_1(\theta) = \lambda_1 Q_{0.75}(q_\theta) + \lambda_2 \, Med(q_\theta) + \lambda^3 E_{q_\theta}[C],$$

where $Q_{0.75}(q_\theta)$ denotes the first quartile (upper quartile), $Med(q_\theta)$ its median, and energy. In practice, each of these quantities is computed from the empirical distribution of $\hat{q}_\theta^{(N)}$. This objective therefore defines a functional $f_1 : \Theta \to \mathbb{R}$ that depends on $\theta$ only through the induced cost distribution $q_\theta$.

We investigate two regimes of aggregation weights. In the first regime, a strongly unbalanced objective is considered, $with \, \lambda_1 = 10^8, \lambda_2 = 10^4, and \, \lambda_3 = 1$. This choice heavily penalizes large upper quantiles of $q_\theta$ and therefore suppresses distributions with heavy upper tails, even if such distributions assign non-zero probability mass to low or optimal cost values. We present in Table 3 the different wavefunctions (a partial description) that are representative of how the wavefunction evolves over the course of the iterations (the numbers in parentheses indicate the probability associated with each cost value).

**Table 3: Probability distribution left shifted for a given instance**

| Cost ↓ | Wavefunction 1 | Wavefunction 2 | Wavefunction 3 | Wavefunction 4 | Wavefunction 5 | Wavefunction 6 |
|---|---|---|---|---|---|---|
| 6 | 0.0 | 1.0 | 3.5 | 2.0 | 3.5 | 0.0 |
| 7 | 0.5 | 3.5 | 0.5 | 0.0 | 0.0 | 0.0 |
| 8 | 0.0 | 0.0 | 14.5 | 15 | 12.0 | 32.0 |
| 9 | 0.0 | 0.0 | 0.5 | 3.0 | 0.5 | 0.0 |
| 10 | 5.0 | 0.0 | 0.0 | 0.0 | 0.0 | 0.0 |
| 11 | 0.0 | 1.5 | 16.5 | 11 | 10.0 | 5.5 |
| 12 | 15.5 | 28.5 | 0.0 | 0.0 | 13.0 | 0.5 |
| 13 | 1.0 | 19.0 | 14 | 6.0 | 0.0 | 24.0 |
| 14 | 0.0 | 1.0 | 0.0 | 1.5 | 1.5 | 0.0 |
| 15 | 7.5 | 0.0 | 0.0 | 0.0 | 0.0 | 0.0 |
| 16 | 1.0 | 0.0 | 0.0 | 0.0 | 0.0 | 10.5 |
| 17 | 0.0 | 0.0 | 0.0 | 0.0 | 0.0 | 0.0 |
| 18 | 0.0 | 0.0 | 0.0 | 0.0 | 0.0 | 21.5 |
| $f_1(\theta)$ | 1200223346 | 1000123317 | 800183000 | 800133421 | 800133313 | 800132631 |

The probability distribution
6.0 ( 3.5) 8.0 ( 12.0 ) 9.0 ( 0.5 ) 11.0 ( 10.0 ) 13.0 ( 13.0 ) 14.0 ( 1.5 )
is associated with a cost of 800133313, but it contains a non-zero probability of finding a solution with cost 6.0.

The probability distribution
8.0 ( 32.0 ) 11.0 ( 5.5 ) 12.0 ( 0.5 ) 13.0 ( 24.0 ) 16.0 ( 10.5 ) 18.0 ( 21.5 ): 800132631
has a lower cost, but a probability of 32% is associated with cost 8.

The second wavefunction is considered better than the first; however, it defines a wavefunction with a lower probability of finding solutions with cost 6, 7, or 5. This example illustrates the significant impact that the objective function will have on the search space exploration. In a second experiment, it could be possible to introduce a balanced objective is used with $\lambda_1 = \lambda_2 = \lambda_3 = 1$, enforcing an even trade-off between tail behavior, central tendency, and mean cost defining a different type of search space exploration.

### 4.3. Remarks on QAOA warm-start

Warm-start strategies admit a natural interpretation in this framework. A warm-start corresponds to selecting an initial parameter vector $\theta_0$ that permit to control how the associated bitstring distribution $p_{\{\theta_0\}}$ is concentrates near bitstrings $x$ with low cost $C(x)$. In favorable cases, subsequent optimization can drive $q_\theta$ toward a near-degenerate distribution concentrated at the optimum, but initial distribution overly concentrated could not favor the search space investigation. Alignment (or misalignment) between initial state and the mixer is difficult challenging problem.

**Example.** Considering the initial solution obtained by removing edges 1–2, 5–3, and 4–5, it is then straightforward to include in the population one or more solutions with different, small angle values, thereby generating candidate solutions centered around the configuration whose cost corresponds to the removal of edges 1–2, 5–3, and 4–5. In this very specific situation, considering $f_1$ we have obtained the following evolution of the wave:

```
6.0   ( 1.0 )  11.0   ( 88.5 )  13.0   ( 3.0 )  14.0   ( 4.0 )  16.0   ( 0.5 )  23.0   ( 3.0 )
6.0   ( 8.0 )   8.0   ( 3.5 )    9.0   ( 2.5 )  11.0   ( 13.5 ) 13.0   ( 3.5 )  14.0   ( 4.0 )
6.0   ( 13.5 )  8.0   ( 5.0 )    9.0   ( 4.5 )  11.0   ( 18.5 ) 13.0   ( 6.5 )  14.0   ( 4.5 )
6.0   ( 35.0 )  8.0   ( 1.0 )    9.0   ( 0.5 )  11.0   ( 24.5 ) 13.0   ( 0.5 )  14.0   ( 2.0 )
6.0   ( 11.5 )  8.0   ( 5.0 )    9.0   ( 3.0 )  11.0   ( 16.0 ) 13.0   ( 3.0 )  14.0   ( 5.5 )
6.0   ( 14.0 )  8.0   ( 8.0 )    9.0   ( 6.0 )  11.0   ( 17.0 ) 12.0   ( 1.0 )  13.0   ( 5.0 )
6.0   ( 27.5 )  8.0   ( 6.0 )   11.0   ( 17.5 ) 13.0   ( 7.5 )  16.0   ( 24.0 ) 18.0   ( 9.0 )
6.0   ( 37.5 )  8.0   ( 3.54 )   9.0   ( 2.0 )  11.0   ( 16.5 ) 13.0   ( 2.5 )  14.0   ( 1.0 )
6.0   ( 37.5 )  9.0   ( 0.5 )   11.0   ( 22.0 ) 13.0   ( 0.5 )  16.0   ( 37.0 ) 18.0   ( 0.5 )
6.0   ( 36.0 )  8.0   ( 1.0 )    9.0   ( 1.0 )  11.0   ( 24.5 ) 13.0   ( 1.0 )  16.0   ( 31.0 )
6.0   ( 39.5 ) 11.0   ( 24.0 )  13.0   ( 1.0 )  16.0   ( 34.0 ) 18.0   ( 0.5 )  23.0   ( 0.5 )
6.0   ( 43.5 )  8.0   ( 0.5 )   11.0   ( 20.5 ) 16.0   ( 33.0 ) 18.0   ( 1.0 )  23.0   ( 0.5 )
```

### 4.4. Concluding remarks

The numerical examples presented in this work are not intended to provide convergence study for QAOA or VQAs but have been introduced to highlight the tight coupling between the definition of the objective function, the choice of classical optimization method, and the initialization strategy, and to show how this coupling determines the structure of the induced cost distribution $q_\theta$. These results feature that optimization quality in variational quantum algorithms must be defined consistently with the intended notion of sampling performance, and that different, commonly conflated criteria correspond to fundamentally distinct mathematical optimization problems.

# 5 Conclusion

Building on the research line introduced by Hadfield (2018), we propose a novel ring-structured mixer Hamiltonian that confines the quantum evolution to the subspace of valid cut configurations. Numerical experiments demonstrate consistency with theoretical predictions, and, to the best of our knowledge, this class of mixers is novel. The key elements of an efficient optimization rely on the definition of a mixer that enables superposition only over bit strings representing valid solutions, as well as on an explicit formulation of the wavefunction that is directly linked to the definition of a warm-start–type approach. In the specific case of the min-cut problem, the approach we have proposed makes it possible to solve the min-cut for large-scale graphs using an iterative procedure in which paths are taken into account iteratively in "batches" of paths.

**Remark:** Author(s) have used Chatgpt to improve English.